\documentclass[sigconf]{acmart}

\usepackage{acmart-taps}
\usepackage{framed}
\definecolor{shadecolor}{rgb}{0.851,0.851,0.851}
\usepackage{listings}
\usepackage{makecell, multirow}
\usepackage{hyperref}
\usepackage{fontawesome}
\usepackage[T1]{fontenc}
\usepackage{xcolor}
\usepackage{colortbl}

\NewDocumentEnvironment{tcolorbox}{ +b }{
    \vspace{10pt}
    \setlength{\arrayrulewidth}{0pt}
    \setlength{\tabcolsep}{1em}
    \noindent\begin{tabular}{|>{\columncolor{white!85!black}}c|}
    \hline
    \parbox{\dimexpr(\linewidth -2\arrayrulewidth -2\tabcolsep )}{%
    \vspace*{2ex}#1\\[0ex]
    } \\ \hline
    \end{tabular}
}

\AtBeginDocument{%
  \providecommand\BibTeX{{%
    \normalfont B\kern-0.5em{\scshape i\kern-0.25em b}\kern-0.8em\TeX}}}

\copyrightyear{2025}
\acmYear{2025}
\setcopyright{acmlicensed}\acmConference[CHI '25]{CHI Conference on Human Factors in Computing Systems}{April 26-May 1, 2025}{Yokohama, Japan}
\acmBooktitle{CHI Conference on Human Factors in Computing Systems (CHI '25), April 26-May 1, 2025, Yokohama, Japan}
\acmDOI{10.1145/3706598.3714064}
\acmISBN{979-8-4007-1394-1/25/04}

\begin{document}

\title{Manifesting Architectural Subspaces with Two Mobile Robotic Partitions to Facilitate Spontaneous Office Meetings}


\author{Ozan Balci}
\email{ozan.balci@kuleuven.be}
\orcid{0000-0002-9265-2262}
\affiliation{
  \institution{Research[x]Design, Dept. of Architecture, KU Leuven}
  \city{Leuven}
  \country{Belgium}
}

\author{Stien Poncelet}
\email{stien.poncelet@kuleuven.be}
\orcid{0000-0003-3248-6781}
\affiliation{
  \institution{Centre for Environment and Health - Dept. of Public Health and Primary Care, KU Leuven}
  \city{Leuven}
  \country{Belgium}
}

\author{Alex Binh Vinh Duc Nguyen}
\authornote{Both authors contributed equally to this research.}
\email{alex.nguyen@kuleuven.be}
\orcid{0000-0001-5026-474X}
\affiliation{
  \institution{Research[x]Design, Dept. of Architecture, KU Leuven}
  \city{Leuven}
  \country{Belgium}
}

\author{Andrew Vande Moere}
\authornotemark[1]
\email{andrew.vandemoere@kuleuven.be}
\orcid{0000-0002-0085-4941}
\affiliation{
  \institution{Research[x]Design, Dept. of Architecture, KU Leuven}
  \streetaddress{Kasteelpark Arenberg 1 - box 2431}
  \city{Leuven}
  \country{Belgium}
  \postcode{3001}
}


\begin{abstract}
Although intended to foster spontaneous interactions among workers, a typical open-plan office layout cannot mitigate visual, acoustic, or privacy-related distractions that originate from unplanned meetings. 
As office workers often refrain from tackling these issues by manually demarcating or physically relocating to a more suitable subspace that is enclosed by movable partitions, we hypothesise that these subspaces could instead be robotically manifested. This study therefore evaluated the perceived impact of two mobile robotic partitions that were wizarded to jointly manifest an enclosed subspace, to: 1) either `mitigate' or `intervene' in the distractions caused by spontaneous face-to-face or remote meetings; or 2) either `gesturally' or `spatially' nudge a distraction-causing worker to relocate.
Our findings suggest how robotic furniture should interact with office workers with and through transient space, and autonomously balance the distractions not only for each individual worker but also for multiple workers sharing the same workspace.  
\end{abstract}



\begin{teaserfigure}
  \includegraphics[width=\textwidth]{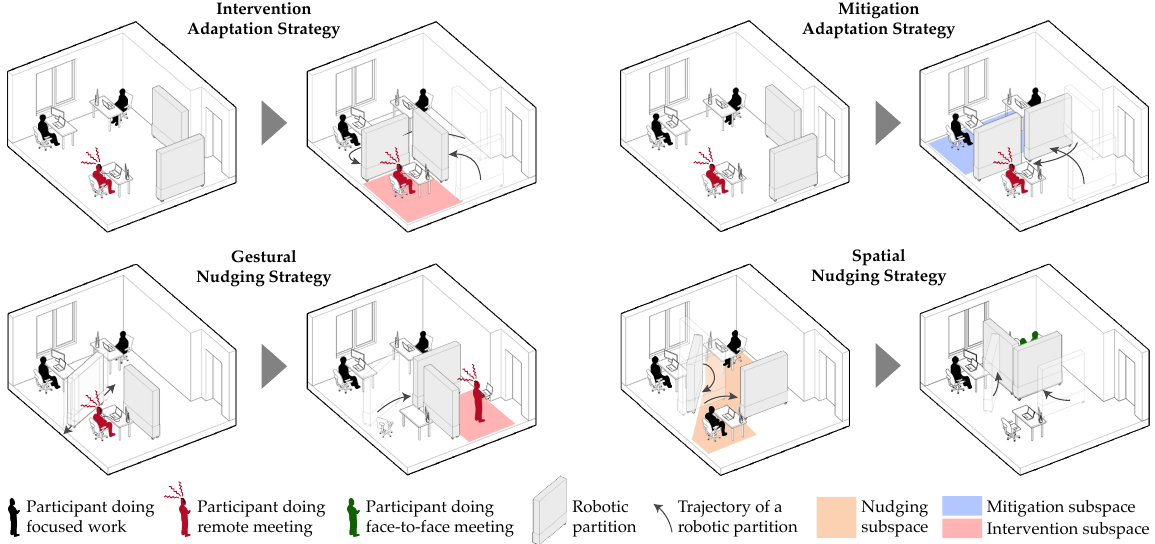}
  \caption{
  To mitigate the distractions originating from a spontaneous meeting, our two wizarded mobile robotic partitions are able to demarcate an open-plan office layout into one or more enclosed subspaces via two adaptation strategies (top) or two nudging strategies (bottom). An adaptation strategy manifested an enclosed subspace either around the source of the distraction, i.e. the participant(s) having a spontaneous meeting; or around the recipient of the distraction, i.e. the participant(s) performing focused work. 
  A nudging strategy either manifested a transient subspace or performed gestural motions to encourage a particular participant to relocate to a different subspace that was more suitable for holding a face-to-face or remote meeting. 
  }
  \Description{
  The adaptation strategies are illustrated through a two-phase sequence depicting how the partitions mitigate distractions arising from spontaneous meetings, through 'intervention', manifesting a subspace around the participant having a spontaneous meeting, and 'mitigation', establishing a subspace for the multiple participants performing focused work. Also, the nudging strategies are presented as a two-phase sequence showing how the partitions guide participants towards another subspace via either 'gestural' nudging, one partition inviting the participant who has a remote meeting towards the direction of the unoccupied subspace, and 'spatial' nudging, both partitions jointly inviting participants to walk through the subspace that they jointly demarcated to have a face-to-face meeting.
  }
  \label{fig:teaser}
\end{teaserfigure}


\maketitle

\section{Introduction}\label{sec:intro}
The knowledge-based industry has adopted the open-plan office layout because it is claimed to accommodate multiple independent workflows, flexible work schedules, and hybrid work arrangements \cite{Kim2016}. Characterised by the absence of physical barriers like walls or cubicles, its spatial openness is meant to encourage spontaneous face-to-face (F2F) interactions among office workers, which in turn supports informal idea-sharing and problem-solving activities \cite{Kaarlela-Tuomaala2009, Steps2019}, enhances creativity and innovation \cite{Samani2020}, and helps build interpersonal relationships \cite{Kim2013} among workers. 
However, environmental psychology research shows that the majority of office workers are dissatisfied with the open-plan layout, as it is unable to mitigate a wide range of acoustic \cite{James2021}, visual \cite{Liebl2012}, and privacy-related \cite{Kaarlela-Tuomaala2009} distractions, many of which are caused by the spontaneous meetings it should encourage. 
As these distractions discourage workers from coming to the office, their absence \cite{Appel-Meulenbroek2022} not only endangers the cohesion and resilience of work teams but also the social well-being of individual workers \cite{Kohn2023}. 
The open-plan office layout therefore requires new solutions that allow spontaneous meetings and focused work activities to coexist.

The so-called `activity-based office' layout, meanwhile, provides multiple enclosed `\textit{subspaces}', each of which is dedicated to optimally host one particular work activity such as meeting, focused work, or relaxation \cite{Gaudiino2023}. 
Yet, office workers tend to be dissatisfied with this layout as frequent switching between different subspaces disrupts their workflow \cite{Hoendervanger2016}, distracts colleagues nearby \cite{Hongisto2016}, and limits their personal agency to customise an individual workstation \cite{Halldorsson2021}.
The so-called `flexible office' layout in turn consists of purposefully designed office furniture elements like mobile desks, partitions, or cupboards that can be manually moved around. Although this layout is meant to empower office workers in spontaneously demarcating smaller subspaces that can better support unplanned work activities \cite{Engelen2019}, workers are required to constantly reflect and then manually adjust their shared workspace. Because these manual adjustments are often repetitive, physically demanding, and require negotiations between co-located workers, the flexible office layout can lead to additional workload or even social embarrassment \cite{Zamani2019}. 
While workers can benefit from dedicated activity-based subspaces, these findings yet suggest that such subspaces should not require additional physical, mental or social effort, and should not disturb colleagues. 

Our study is based on the premise that dedicated activity-based subspaces could become autonomously manifested by multiple mobile robotic partitions. This premise has become technologically feasible, as robotic furniture is now able to manoeuvre life-sized architectural elements like sofas \cite{Spadafora2016}, work desks \cite{Kim2021}, office chairs \cite{Agnihotri2019}, or trash cans \cite{brown2024trash} inside human-inhabited environments, such as shared offices \cite{takashima2015movementable}, public buildings \cite{Agnihotri2019}, and even urban environments \cite{brown2024trash}.
Moreover, robotic furniture has been shown to be able to convey specific intentions to co-located humans, insofar as a robotic partition changed its form to encourage individuals to approach \cite{Lee2013}, perpetually moving robotic chairs nudged passers-by to sit down \cite{Agnihotri2019}, and moving tables enticed people to collaborate with one another \cite{takashima2015movementable}.

In the context of adaptive architecture, i.e. the life-sized integration of robotic technology that alters architectural layout, a single mobile robotic partition was able to spatially separate distinct activities \cite{Nguyen2022}, reduce some spatial distractions \cite{Nguyen2024}, or improve the privacy of individual workers \cite{Onishi2022} in shared workspaces.
Moreover, people were able to interpret the static manifestation and dynamic motion of one single robotic partition to the extent that they felt nudged to change their work activities \cite{Nguyen2022} or relate their mutual visibility to their social availability \cite{Nguyen2024}. 
Based on these promising findings, we deepen the premise of past studies by aiming to robotically manifest a fully enclosed subspace instead of only demarcating one spatial boundary, and to nudge workers to move to this enclosed subspace instead of changing their work activity. 
As such, this study envisions a future of adaptive architecture that allows an open-plan layout to autonomously reduce office distractions by merging the advantages of the flexible and the activity-based office layout. However, it is still unknown how such an enclosed subspace should be manifested, and how a worker causing distractions should be encouraged to use it.

Our primary research question is therefore: `\textbf{How can two autonomous robotic partitions manifest an enclosed subspace within an open-plan office layout to reduce distractions originating from a spontaneous meeting?}'.
Using the Wizard-of-Oz method \cite{Riek2012} to control two seemingly autonomous mobile robotic partitions, we deliberately deployed them in a relatively small, existing open-plan workspace so that an enclosed subspace could be manifested by four orthogonally adjoining vertical boundaries, consisting of two adjoining physical walls and our two robotic partitions.
We invited nine groups of three participants each to concurrently engage in either a spontaneous meeting (i.e. remote or F2F) or a focused work activity. We deliberately deployed the study in an existing office building and invited participants who regularly work in an open-plan office layout, many of whom worked in the same building, to enhance the ecological validity.

As illustrated by \autoref{fig:teaser}, our study exposed participants to four different architectural strategies, each manifesting a dedicated subspace to reduce the visual, acoustic and privacy distractions of the spontaneous meeting.
The two \textit{adaptation strategies} are grounded on how office workers interpret a subspace that separates the source of a distraction differently from a subspace that protects the recipient of a distraction in terms of its spatial and social impacts \cite{Nguyen2024}. As such, we contrast a subspace that encloses a worker conducting a remote meeting with a subspace that encloses others who feel distracted by the acoustic, visual and privacy consequences of this remote meeting.    
The two \textit{nudging strategies} were designed to encourage the workers in a spontaneous meeting to relocate to a more suitable, newly manifested subspace. As architectural designers, we grounded these two nudging strategies on the hypothesis that while robots can meaningfully interact with humans via mostly motion-based gestures \cite{Fink2014, ju2009approachability, sirkin2015mechanical}, there still exists an untapped potential to indirectly guide people by manifesting a transient space. As such, we contrast the perpetual motion of robotic partitions with the architecturally dramatic opening and closing of a subspace that visibly connects two workers.

Through a qualitative analysis of self-reported questionnaires, group interviews \cite{frey1991group}, and observations, we discovered how co-located workers preferred contradicting adaptation strategies based on their differing work activities, and preferred the indirect yet still intuitively understandable spatial nudging strategy as it was considered less intrusive.
By reflecting on these discoveries in relation to our goal to reduce open-plan office distractions, we discuss 3 design considerations: 
(1) A newly manifested subspace introduces new distractions by how its actual static, morphological and dynamic qualities misalign with the expectations of the workers engaging in the work activity it hosts;
(2) These newly introduced distractions should be balanced with the reduced distractions, on the level of one subspace that supports one particular work activity, but also on the level of multiple subspaces that coexist in the same workspace;
(3) Next to creating subspaces around workers, robotic partitions can nudge workers by manifesting a transient subspace that regulates the visual and physical access among workers, as long as it is subtle and non-distracting.
We believe these insights could be applied to other types of robotic furniture or even to other robots operating in human-inhabited environments. Additionally, they highlight the importance of designing mobile robots to better leverage their architectural affordances.

\begin{figure*}[t]
    \centering
    \includegraphics[width=1\linewidth]{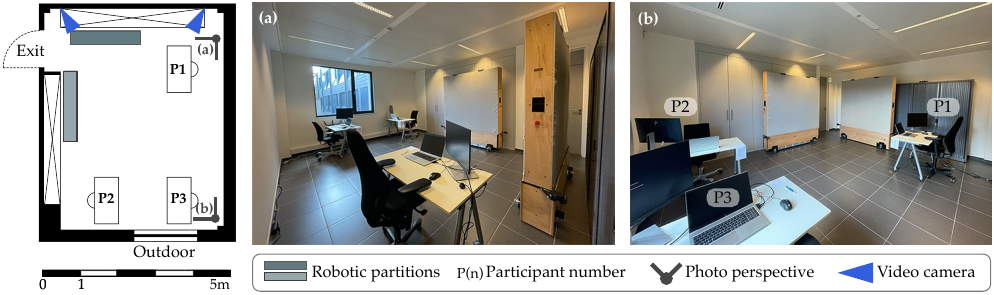}
    \caption{(Left) The open-plan layout of the workspace in the baseline condition, featuring three separate office desks, one window and one door.
    Two strategically positioned cameras captured the behaviour of the participants throughout the study. (Right) Photos of the workspace showcasing the open-plan layout without any robotically manifested subspace.
    }
    \Description{Left: A plan of the open office featuring one window facing southeast, one exit door and three separated desks: P1 facing the exit door, P2 and P3 facing each other next to the window.  Two laptops with cameras were positioned on the top of the free-standing closet to video record the experiment. The two robotic partitions are positioned in their baseline layout. Right: Two photos of the open office, showing a modest postmodern interior design that is characterised by a high ceiling, one built-in cupboard, one free-standing cupboard and a tiled floor.
    }
    \label{fig:studySpace}
\end{figure*}

\section{Related Work}
We base our study on previous findings on the perceived distractions in open-plan office layouts, the psychological impact of changing office layouts, and how people interpret the autonomous behaviour of robotic furniture.

\subsection{Open-plan Office Layout Distractions} 
The main dissatisfaction with open-plan offices arises from acoustic and visual distractions and a lack of privacy, often due to spontaneous meetings \cite{Kaarlela-Tuomaala2009}. This is critical to address as most office workers spend the majority of their time on focused tasks requiring concentration \cite{Newsham2005}.

While continuous background noise is less bothersome, sudden and comprehensible interactions, such as spontaneous, informal coworker meetings are the primary distractions in open-plan offices that hinder concentration \cite{Felipe2023}. Visual distractions caused by co-workers' activities such as using office facilities, performing physical tasks, or walking in the office \cite{Hongisto2016, Liebl2012, Berti2001}, also divert worker attention \cite{Nicholls2005}.
When workers cannot isolate themselves from these acoustic and visual distractions, they feel a lack of control over their environment, which even amplifies the impact of distractions and diminishes their sense of privacy \cite{Bodin2009}. 

The open-plan office layout also decreases the level of control workers have over their social interactions and interruptions, termed \textit{task privacy}, which is itself a source of distraction \cite{Oldham1988}. The diminished personal enclosure exposes workers to unwanted observation, compromising their sense of \textit{visual privacy} and perceived personal control over their workspace \cite{Brand2005, Danielsson2009}. Additionally, \textit{acoustic privacy}, or the ability to hold confidential conversations and safeguard sensitive information, is often diminished \cite{Weber2019}.

To reduce these distractions and respond to this lack of control, workers often manually adapt their environment by themselves. This involves using furniture elements, such as shaders, blinders or partitions to create visual or acoustic barriers \cite{Roskams2021c}. Workers may also choose to relocate to another area \cite{Gaudiino2023} or move the source of distraction to another part of the office. However, these manual adaptations are often deemed as being insufficient as they cannot reduce the distractions as desired, or lead to introducing distractions to others, such as by moving furniture within the field of view of others \cite{Hongisto2016, Appel2017}. The presence of others in the open-plan office layout tends to discourage workers from making adaptations, as they may feel less empowered to prioritise their own needs and assume others will handle the distraction \cite{Reinhart2003}. In such cases, workers with greater access to resources for performing adaptations are more likely to act \cite{Day2012}.

\subsection{Impact of Changing Office Layouts}
As humans possess a natural psychological tendency to evaluate their past and present circumstances after a significant change, workers experience a certain degree of discomfort when their office layout changes.
This inclination to assess change often heightens awareness of both gains and losses between the original and new office layouts \cite{Morrison2021}, in which established theories of loss aversion \cite{Kahneman1991} suggest that individuals are more sensitive to perceived losses than to gains. For instance, workers experience a pronounced loss of privacy when moving from a closed to an open-plan office setting \cite{Morrison2021}, as they suddenly have to share their workspace with others. Emotional attachment to the previous workspace can also make it difficult to adjust, as workers struggle with the loss of familiar surroundings and a diminished sense of identity in the new layout \cite{Neuner2006}. Yet, despite these challenges, the perceptions of workers on a new office layout are often influenced by how well it meets their needs in providing support (e.g. access to daylight, colleagues, meeting rooms, etc.) to retain motivation and productivity \cite{Cobaleda2020}. 

\subsection{Robotic Furniture}
Expanding upon the concepts of `robotic building' \cite{bier2016robotic} and `architectural robotics' \cite{green2016ecosystems}, the domain of robotic furniture \cite{sirkin2015mechanical, ju2009approachability, sadka2022way, fink2014robot} proposes the integration of sophisticated robotic technology into everyday human life \cite{di2011peis}. Recent advancements in this domain showcase that robotic furniture can now not only react to the input or needs of occupants \cite{gronbaek2017proxemic} at home \cite{zhong2023exploring} or at work \cite{Onishi2022}, but also `nudge' their behaviour \cite{Caraban2019} towards ergonomic \cite{fujita2021tiltchair} or collaborative \cite{takashima2015movementable} purposes.

Certain robotic furniture employed subtle, almost unnoticeable motions to nudge people in the background of their attention, like a sit-stand desk that slowly adjusts its height \cite{lee2019effects, Kim2021}, a chair that moderately inclines its seat \cite{fujita2021tiltchair}, or a workstation that gradually adjusts its desk, chair, and computer monitor \cite{wu2018activeergo} to enhance occupants’ posture ergonomically during work activities.
In contrast, other robotic furniture employed prominently noticeable motions to explicitly nudge people, such as a robotic footstool prompting people to rest their feet by moving towards them \cite{sirkin2015mechanical}, a pair of robotic chairs encouraging passers-by to sit down and play chess by moving back and forth \cite{Agnihotri2019}, a robotic door inviting passers-by to enter by repeating its `waving' open-and-close motion \cite{ju2009approachability}, or a pair of robotic bar stools motivating seated people to interact by rotating them towards each other \cite{sadka2022way}. 

Previous studies on robotic partitions have shown how they not only could react to the needs of occupants, such as by dynamically reconfiguring subspaces based on occupant input \cite{Onishi2022}; but also nudge occupant behaviours, such as by employing physical motions in expanding or compressing a subspace to communicate certain approachability \cite{Lee2013}, or by deliberately modifying a subspace to encourage an occupant to change their activities \cite{Nguyen2022}.


\section{Methodology}

Our semi-controlled study \cite{kircher2017design} was conducted at the local headquarters of IDEWE \footnote{IDEWE: \href{https://www.idewe.be/}{idewe.be}}, which is a consulting company specialising in healthy and safe work environments. This office building was chosen because the company appreciated our research motivations and allowed us to maintain ecological validity by recruiting real knowledge workers with direct experience in open-plan offices.
We selected the relatively small-sized (\(28m^2\)) workspace shown in \autoref{fig:studySpace}, situated on the ground floor with a single door and a single southeast-facing window. 
We purposefully chose the location of each office desk so that the two relatively large robotic partitions could manifest enclosed subspaces without colliding with the office furniture.

\subsection{Technical Implementation}
\begin{figure*}[t]
    \centering
    \includegraphics[width=1\linewidth]{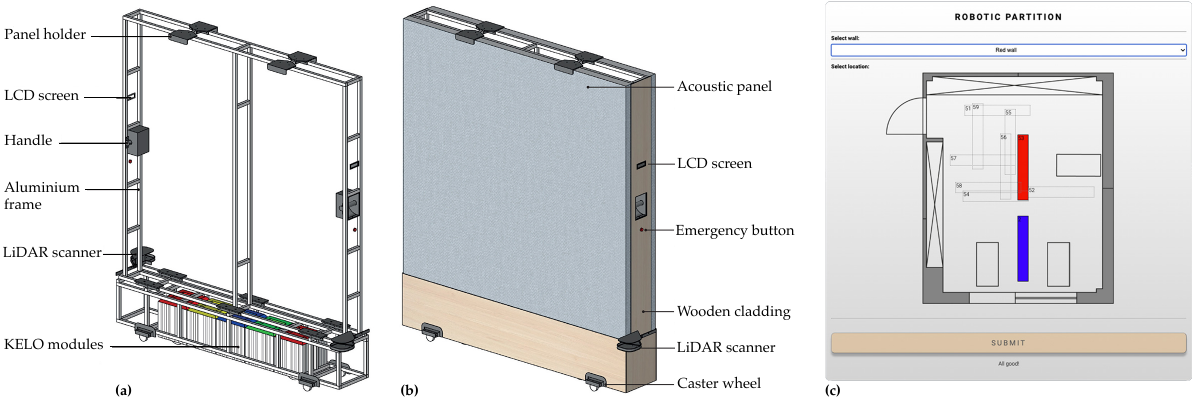}
    \caption{The robotic partitions adopted a similar design to that of Nguyen et al. \cite{Nguyen2024}. (a) Each robotic partition featured a 180*210*28 cm lightweight aluminium frame that houses a customised robotic configuration. 
    (b) The frame carried two acoustic panels and wooden cladding and incorporated handles for manual manoeuvring, emergency stop buttons, and a small LCD screen. Notably, two extrinsically-positioned LiDAR laser scanners were required to ensure a 360-degree field of view around its perimeter. 
    (c) A web-based interface enabled the researcher to select a predefined position, to which each robotic partition could autonomously navigate.
    }
    \Description{Description of the robotic partitions. (left) a 180x210x28cm lightweight aluminium frame that houses a customised configuration of two robotic drive wheels, a battery with a power distribution board, a wireless control unit, and two extrinsically positioned LiDAR laser scanners to ensure a 360-degree field-of-view around the partition’s perimeter. (middle) The partition is covered by two acoustic panels and wooden cladding, which includes handles for manual manoeuvring, emergency stop buttons, and a small LCD screen. (right) The web-based interface shows the selection of a position from a pre-defined list and the transmission of the command to each partition via a cloud.
    }
    \label{fig:partitionTech01}
\end{figure*}

As shown in \autoref{fig:partitionTech01}, the overall morphology of the two identical robotic partitions, characterised by two acoustic panels and wooden cladding, was specifically designed to embody a standard office partition with sound-dampening capabilities.
The silent, reliable and safe (\(0.2\) to \(5 cm/s\)) mobility of each partition was facilitated through the KELO Robile robotic platform \footnote{KELO Robile: \href{https://www.kelo-robotics.com/products/\#kelo-robile}{kelo-robotics.com}}. 
To minimise the depth of a partition, the four industrial-grade KELO modules were assembled on a single linear axis with two drive wheels positioned on opposite ends (see \autoref{fig:partitionTech01}). 
The semi-autonomous mobile capability was achieved via custom-developed software built within the Robot Operating System (ROS) framework.
This software allows each partition to autonomously localise itself and independently navigate to predefined destinations while avoiding obstacles within a \(30 cm\) range.
Two emergency buttons integrated on both sides of the robotic partition allowed users to instantly disable the wheels if required.
The position of each robotic partition was semi-controlled via the Wizard-of-Oz method, as a covert researcher selected the sequential destination of each robot to autonomously navigate to in real-time.

\subsection{Strategies}
The study involved the comparison of four different strategies along with a baseline condition.

\subsubsection{Baseline Condition}
\begin{figure*}[t]
    \centering
    \includegraphics[width=1\linewidth]{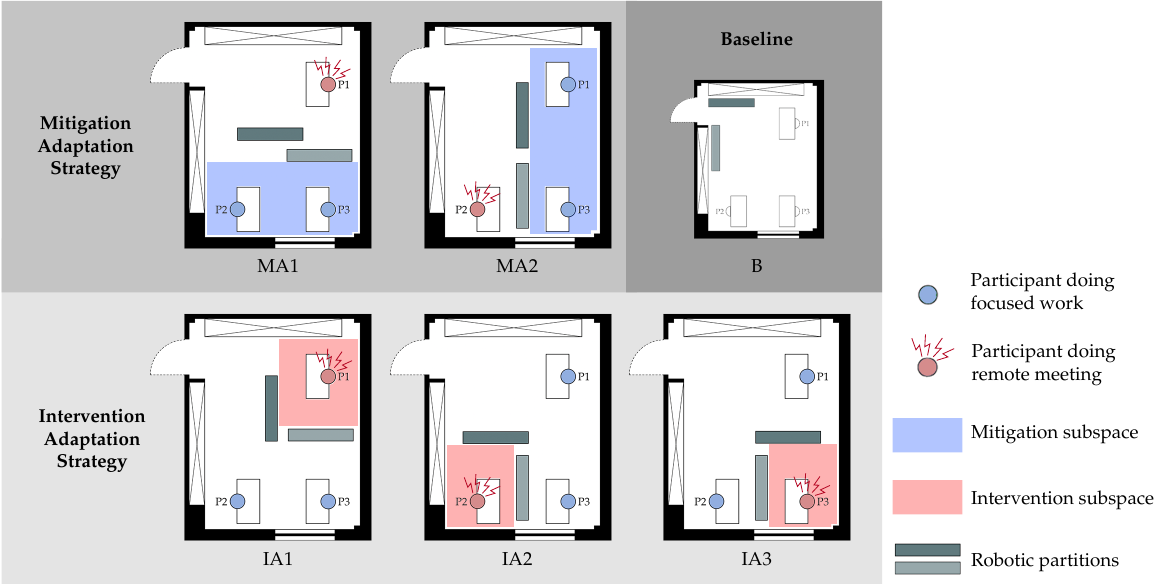}
    \caption{
    The adaptation strategy conditions consisted of one \textit{baseline} (B) condition, whereby no subspace was manifested; three \textit{intervention} (IA1-3) subspaces that manifested around the participant holding a remote meeting; and two \textit{mitigation} (MA1-2) subspaces that manifested around the shared space of the two participants doing focused work activities. 
    One particular mitigation subspace could not be implemented due to the limited space constraints in this workspace.
    }
    \Description{Overview of layouts based on the adaptation strategies. (top-right) Baseline layout, whereby the partitions are positioned out of sight to simulate a typical open-plan office layout. (top-left) Two mitigation subspaces, whereby a subspace is demarcated around the shared social areas of the two participants doing focused work. (bottom) Three intervention subspaces, whereby a subspace is manifested around the participant who is in a remote meeting.
    }
    \label{fig:adaptationStrategies}
\end{figure*}

As shown in \autoref{fig:adaptationStrategies}, the two robotic partitions were positioned along the boundaries of the workspace during the baseline condition so that the layout could serve as a prototypical open-plan layout benchmark against which the adaptation strategies could be compared.

\subsubsection{Adaptation Strategies}
The adaptation strategies were designed around how one worker unexpectedly started a remote meeting on their computer, thereby disturbing two colleagues in focused work. \autoref{fig:adaptationStrategies} depicts the six possible subspaces that can be manifested by the two adaptation strategies.
Accordingly, the `intervention' subspace aims to reduce distractions by enclosing the participant in the remote meeting. In contrast, a `mitigation' subspace encloses the distracted participant(s) doing a focused work activity.
Not only motivated by previous robotic partition studies \cite{Nguyen2024} as outlined in the \nameref{sec:intro}, the two adaptation strategies also draw inspiration from environmental psychology, where workers in open-plan office settings typically employ two primary coping mechanisms to manage environmental demands: approach and avoidance \cite{Roth1986}. The intervention strategy reflects approach coping, which involves actively addressing the source of distractions, while the mitigation strategy aligns with avoidance coping, focusing on diverting attention away from distractions \cite{Skinner2003}. Recognising that workers often choose between these coping mechanisms due to social inhibitors rather than their effectiveness \cite{Fanger1970, Galasiu2006}, we aimed to examine whether their autonomous manifestation through the adaptation strategies would produce similar effects.

We hypothesised that participants interpret the underlying intentions of each adaptation strategy differently, such as whether an adaptation strategy seems to `attack' or `protect' a worker. This hypothesis is supported by other robotic furniture studies, where participants were able to relate specific robotic motions with social intentions \cite{sadka2022way, takashima2015movementable}.
Building on theories of spatial perception and environmental psychology \cite{hall1968, Wicker2002}, we hypothesised that the subspaces manifested by each adaptation strategy would be experienced differently, due to their apparent spatial size, proportionality, and ability to regulate views.

\begin{figure*}[t]
    \centering
    \includegraphics[width=1\linewidth]{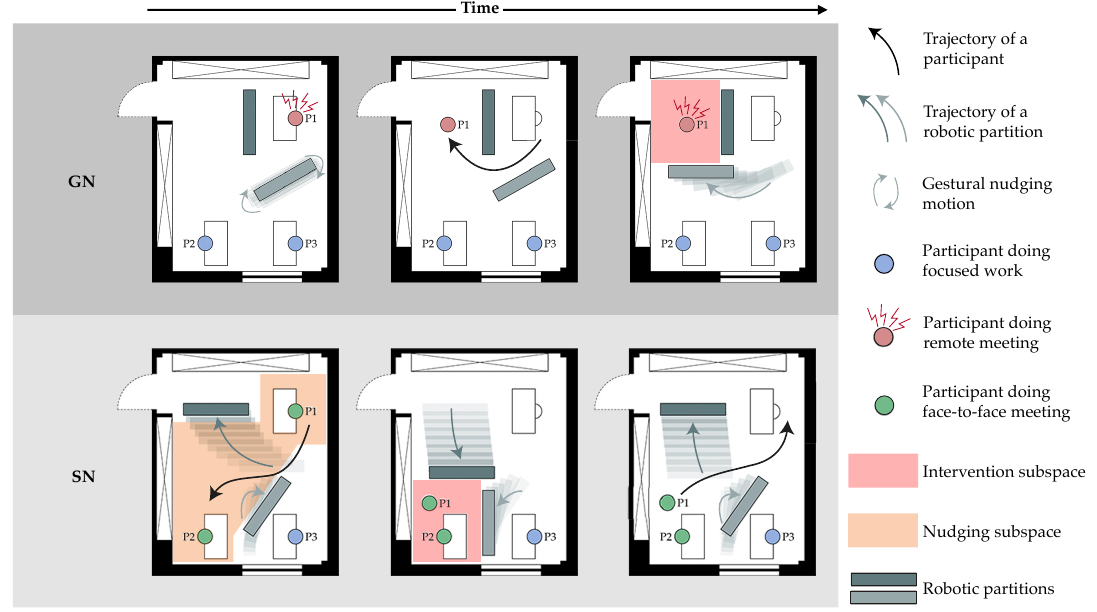}
    \caption{
    (Top) The gestural nudging (GN) strategy consisted of one robotic partition performing a `swaying' motion to encourage the distracting participant engaged in a remote meeting to enter the intervention subspace.
    (Bottom) The spatial nudging (SN) strategy gradually manifested a transient subspace that regulated the visual and physical access between two participants conducting a face-to-face meeting to encourage participants to initiate and conclude the meeting.
    }
    \Description{
    Figure illustrating the trajectories and actions of participants (P1, P2, P3) and robotic partitions in an open-plan office environment over time. The top row represents gestural nudging (GN), where robotic partitions use perpetual motions to nudge occupants, while the bottom row represents spatial nudging (SN), where subspaces manifest subtly through spatial qualities. The figure highlights participants engaged in focused work, remote meetings, and face-to-face meetings, with intervention and nudging subspaces depicted in red and orange, respectively. Key elements include participant movement, partition trajectories, and the transient formation of subspaces to influence occupant behaviour.
    }
    \label{fig:nudging_combined}
\end{figure*}

\subsubsection{Nudging Strategies}
The nudging strategies were designed around initiating two prototypical spontaneous meeting scenarios. The first scenario resembled that of the adaptation scenario, i.e. a single worker starting a remote meeting that is disturbing to their colleagues. The second scenario involves two workers who are spontaneously meeting each other face-to-face, thereby disturbing their colleague doing focused work. 
As shown in \autoref{fig:nudging_combined}, the gestural nudging strategy (GN) involved one partition manifesting an intervention subspace and the other executing a perpetual `swaying' motion. Once a participant entered this intervention subspace, the robotic partition enclosed it.
The spatial nudging strategy (SN) involved both of the partitions manifesting a transient subspace that regulated the physical and visual access between two participants engaged in an F2F meeting. Once one participant entered the newly manifested subspace, both partitions enclosed it. Upon concluding the spontaneous meeting or after 2 minutes, whichever occurred first, the subspace manifested the same transient subspace again to nudge the visiting participant to leave.
We hypothesised that both nudging strategies differ as participants have to interpret the nudging directly by perceiving a motion pattern from the partitions, versus by indirectly experiencing the subspace they jointly manifest.

This hypothesis is based on the premise that occupants can interpret the intentions behind certain robotic motions \cite{hoffman2014}, insofar that they even can be nudged \cite{ju2009approachability}. Similarly, occupants can interpret the implicit architectural meaning of certain subspace configurations \cite{Bolbroe2016}, insofar that people were nudged by directing their views \cite{Alavi2018}, steering their circulations \cite{Natapov2020}, or even suggesting their work activities \cite{Nguyen2025}.

\subsection{Recruitment}
14 of the 27 (\(52\%\)) participants, whose demographics are detailed in Appendix E, were recruited internally from the local company via an email newsletter, whereas 13 (\(48\%\)) participants were recruited externally from our university using email lists and social channels. The recruitment email outlined the study motivation, duration, and location, and offered participants either a 10-euro multimedia store coupon or a charity donation. Interested participants selected available time slots through an online form after confirming they had experience working in an open-plan office layout.
Consequently, three groups (\(33\%\)) consisted of participants already acquainted with each other, while the researchers allocated the remaining 18 unacquainted participants into six groups (\(67\%\)). The final cohort thus included four groups of local company workers, three groups of external participants, and two mixed groups. Recruitment was concluded after 27 participants were confirmed, forming a common participant cohort for qualitative HRI studies \cite{Conlon2020}.

\subsection{Procedure}
\begin{figure*}[t]
    \centering
    \includegraphics[width=1\linewidth]{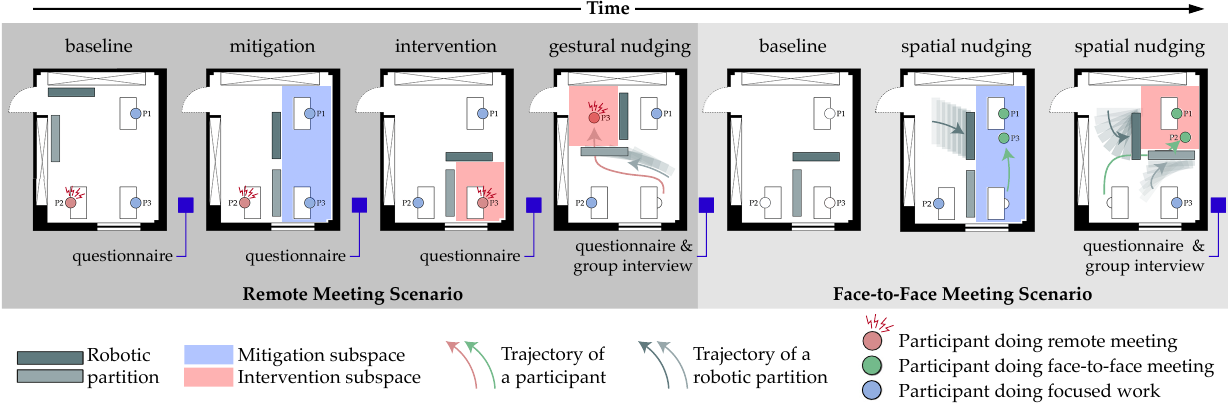}
    \caption{Example of the study procedure as experienced by Group 4 (G4):
    The study began with the remote meeting scenario, starting with the baseline layout, followed by three different strategies. It then progressed to the face-to-face meeting scenario, which also began with a baseline layout where participants had no visual or physical access to one another, followed by two spatial nudging strategies.
    }
    \Description{Example of the seven-phase sequence of applied layouts based on the adaptation and nudging strategies during the experiment of group 4. The first four phases belong to the remote meeting scenario and the other three phases belong to the face-to-face meeting scenario: (1) baseline layout (partitions are out of sight to create an open-plan office) when participant 2 is in a remote meeting; (2) mitigation strategy when participant 2 is in a remote meeting; (3) intervention strategy when participant 3 is in a remote meeting; (4) gestural nudging strategy when participant 3 is in a remote meeting; (5) baseline layout whereby the partitions are positioned in an L-shape between the participants; (6) the partitions spatially nudge participant 3 to have a face-to-face meeting with participant 1; (7) the partitions create an intervention subspace around the desk of participant 1, while participant 1 and 2 are having a face-to-face meeting. During the remote meeting scenario, each strategy was alternated with a questionnaire to reflect on the experienced layout, ending with a group discussion. The face-to-face meeting scenario was evaluated and concluded with a questionnaire and group interview.
    }
    \label{fig:studyTimeline}
\end{figure*}

As shown in \autoref{fig:studyTimeline}, each participant group started with a familiarisation stage, followed by experiencing two adaptation strategies and two nudging strategies.
The order of the conditions for each group is detailed in Appendix B, which shows how each pair of strategies was purposefully ordered across the nine experiments, ensuring that the sequence in which one strategy occurred relative to another was evenly balanced throughout the study.
The apparent autonomous behaviour of both robotic partitions was wizarded by a researcher observing the workspace from the neighbouring hallway via two live camera feeds. A second researcher was either inside the workspace to give verbal instructions and answer questions, or outside to act as the second attendee in the remote meeting.

\subsubsection{Familiarisation Session}
Participants were first introduced to the purpose of the study, which aimed to investigate how robotic partitions can autonomously manifest subspaces to reduce spontaneous meeting distractions in an open-plan office layout.
After signing an informed consent form, participants observed the motions of the robotic partitions, experiencing their seemingly autonomous capabilities first-hand.
They also learned the purpose of the `swaying’ motion used in the gestural nudging strategy, which encouraged them to enter another subspace.

\subsubsection{Remote Meeting Scenario} 
In this scenario, two participants engaged in focused work by completing a calculation sheet, while one participant engaged in a remote meeting with the second researcher. The distractions of this remote meeting were meant to be reduced by manifesting either a mitigation or an intervention subspace, the last of which was augmented with the gestural nudging strategy. 
To ensure that they could compare two different strategies under a similar work context, participants only altered their work activity after experiencing two consecutive strategies.

After each strategy, all participants completed a Likert-scale questionnaire that focused on how they experienced: 1) their work activity; 2) the workspace atmosphere; 3) the impact of the subspace on their work activities; and 4) the dynamic behaviour of the robotic partitions. Participants were also asked to reflect more qualitatively on all these questionnaire responses by answering an open-ended question. After all four strategies, the two researchers conducted a 10-minute semi-structured group interview \cite{frey1991group} with the participants, focusing on their overall experience. The questionnaire results were employed during the group interviews, guiding discussions on the different experiences between individual participants.

\subsubsection{Face-to-face Meeting Scenario}
In this scenario, two predetermined participants were informed that they should have an F2F meeting with the remaining participant. An F2F meeting involved a short brainstorming session on a given topic, with responses typed on one participant’s laptop.
Participants were not told when to approach another participant but were instructed to interpret the transient subspace manifested by robotic partitions.
The distractions of this F2F meeting were meant to be reduced by manifesting either a mitigation or an intervention subspace.
As shown in Appendix B, the pairs of participants to hold an F2F meeting were pseudo-randomised. 

After two subsequent F2F scenarios with different pairs of participants, all participants were invited to complete a Likert-scale questionnaire about how they experienced: 1) the overall atmosphere; 2) the dynamic behaviour of the robotic partitions; 3) the impact of the transient and static subspaces on the meeting; and 4) whether it effectively nudged them to initiate or finish their F2F meeting. Finally, the researchers conducted a 10-minute semi-structured group interview \cite{frey1991group}, focusing on the spatial nudging strategy and their overall experience.

\subsection{Data Acquisition and Analysis} 
The collected data included participant responses from the questionnaires; video recordings from two video cameras; audio recordings from the semi-structured group interviews; and observations of noteworthy events made by the researchers.

The questionnaire responses were analysed to reveal any prominent differences between the strategies, while the video recordings and observations proved helpful in tracing back participant behaviours.
The audio recording was transcribed, before a thematic analysis \cite{Braun2006} was conducted to identify, analyse and report patterns of participant experience, in triangulation with the questionnaire responses and the video recordings. One author carried out the initial coding to identify patterns and generate preliminary categories. Codes were assigned by analytical memo writing and discussions among all authors. From the 36 codes that resulted, 11 categories of abstract patterns were identified. These categories were treated as working hypotheses to be iteratively merged and split through group discussions between the four co-authors using a constant comparative method. This process resulted in 21 dimensions, categorised across the three different work activities and the four adaptation or nudging strategies, as reported in the \nameref{sec:results}.


\section{Results}\label{sec:results}
We recruited 27 participants, consisting of 18 women and 9 men, aged from 18 to 64 (\(Mean=35.72, SD=11.0\) years), most of whom (\(n=15,55.6\%\)) had more than five years of experience working in an open office layout. Each group of participants took part in the study for a duration ranging from 64 and 86 minutes (\(Mean=75, SD=7.5\) minutes). In this section, each participant (e.g. G1P2) is coded by their group number (e.g. G1) and their desk location (e.g. P2, as shown in \autoref{fig:studySpace}). The number \((n)\) following each code represents the number of groups in which it was mentioned and discussed, as we assume that uncontested utterances from individual participants were unanimously accepted by all other group members. 

\subsection{Adaptation Strategies}
Knowing that the type of work activity influences how workers perceive their environment \cite{vandenBerg2020}, we subdivided our results by work activity, and then how the adaptation strategies reduced or introduced certain distractions.

\subsubsection{Focused Work}\label{sec:adaptationfocused}
Participants engaged in focused work placed particular importance on maintaining visual access to other focused participants, as well as to the window and the door.

\begin{figure*}[t]
    \centering
    \includegraphics[width=1\linewidth]{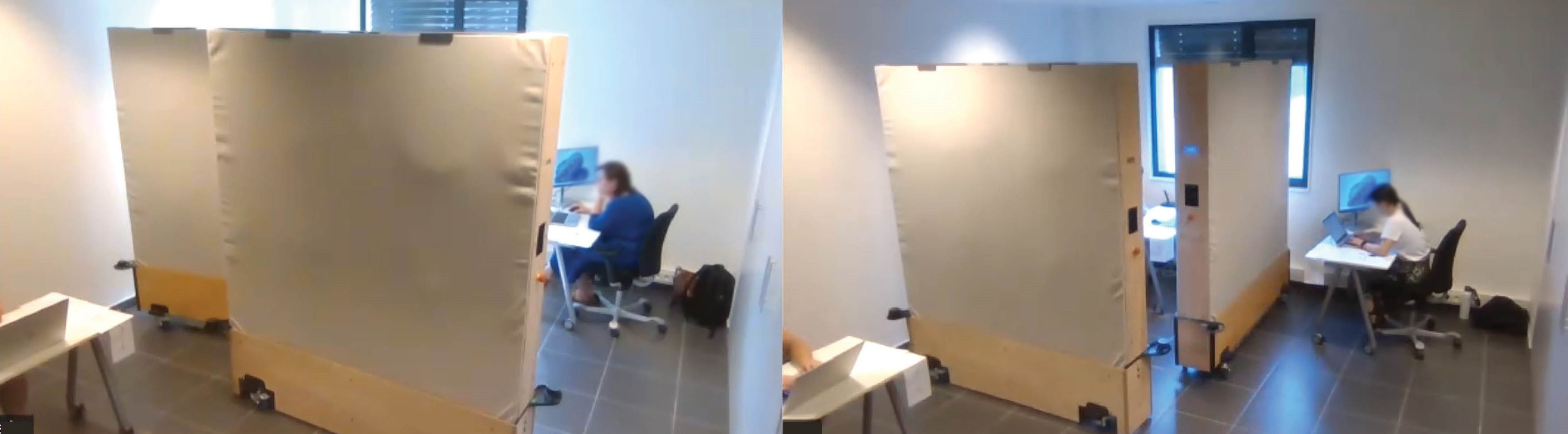}
    \caption{During focused work, a P3 participant was distracted by how the partitions of the mitigation subspace (MA1-left) were misaligned with each other, while an intervention subspace for P3 (IA3-right) covered the window view of P1 and P2. 
    }
    \Description{Two photos of how the partitions affected participants doing focused work. (left) The partitions are positioned in the mitigation subspace (MA1) and are not aligned. (right) The partitions are positioned in the intervention subspace (IA3), covering the window from  P1 and P2.}
    \label{fig:banner4.1.1}
\end{figure*}

\paragraph{Reducing Distractions} 
Six in nine groups reported the intervention subspace to be more effective in reducing visual distractions from the participant having a remote meeting during focused work: ``\textit{The [partitions] covered my view to him making his activity less distracting}" (G7P3); ``\textit{Because I knew I couldn't see him, I wasn't looking}" (G8P2). However, the mitigation subspace was considered more visually supportive than the intervention subspace, as participants ``\textit{liked [seeing the other participant doing focused work]}" (G9P2) together in the same subspace.
One group, however, judged the mitigation subspace as less effective than the intervention as a small gap between the partitions left the remote meeting participant visible (G3P1-MA1).

Most groups could not perceive a difference in the acoustic dampening between the two adaptation strategies. Six groups noted that while ``\textit{[the partitions] give a little bit less noise}", it was ``\textit{not sufficient to become actually concentrated}" (G4P3); as they were ``\textit{still very distracted}" (G2P1) by their colleague doing a remote meeting. On the other hand, three groups reported perceiving less acoustic distraction when the participant conducting a remote meeting was gesturally nudged to enter the intervention subspace thanks to the further distance (G1P3).
However, seven groups stated that they did not perceive any acoustic differences in the same condition (G5P2).

\paragraph{Introducing Distractions} 
The visibility of the window emerged as a crucial spatial quality for focused work, as four groups critiqued the intervention subspace IA3 (as shown in \autoref{fig:banner4.1.1}) because they: ``\textit{couldn't receive any light from the window}" (G7P2) or ``\textit{didn't see a window anymore, so [...] no outdoor connection}" (G8P1). Three groups (G1, G4, G6) argued that a lack of windows was tolerable for a short period because they were ``\textit{really focused on the task}, and thus were ``\textit{not really looking out of the window.}" (G6P1). 
One group felt uncomfortable with the mitigation subspace (MA2) because it blocked the view to the door, making it ``\textit{claustrophobic}" (G3P3), and felt more at ease with the intervention subspace IA2.

\subsubsection{Remote Meeting}\label{sec:adaptationobtrusive}
Participants engaged in a remote meeting particularly foregrounded how the subspace afforded privacy.

\paragraph{Reducing Distractions} 
Seven groups felt more satisfied with the visual privacy offered by the intervention subspace than the mitigation subspace during a remote meeting, particularly because it ``\textit{made an enclosed, private space to support [their] remote meeting}" (G7P3).
However, this subspace was considered ineffective at reducing acoustic distractions, as a participant in a remote meeting ``\textit{[...] had the privacy visually and [...] thought, 'okay, they are fine [...]'. I am very enthusiastic. And then my voice goes louder and louder. I really think I didn't disturb others}" (G1P2). 
This negative acoustic effect became amplified by wearing headphones, making participants in a remote meeting unaware of the acoustic distractions they created: ``\textit{I was much more focused when I was in the meeting with my headphones that cut off all of the noise.}" (G2P1).
Once two participants realised that the subspace was apparently unable to dampen the acoustic distraction, they ``\textit{[...] worried that [other participants] were distracted just by hearing me}" (G3P2). 

\paragraph{Introducing Distractions}
Some of the participants in remote meetings critiqued the close proximity of the robotic partitions: ``\textit{Personally I don't like facing the [partition]. I feel kind of isolated, annoyed}" (G9P1); as well as because they ``\textit{[...] don’t like staring at the wall. If there is a picture then there is something to look at.}" (G9P1). As the study took place during a relatively hot summer, two groups felt ``\textit{[...] quite warm here, and especially when [in intervention subspace], you do feel it}" (G2P2) as it cut off the ventilation from the window.
Consequently, they found this subspace suitable only for short periods, as prolonged use would cause discomfort: ``\textit{If I need to do it like every day for 8 hours, for example, and the [partitions] are like very much enclosing me, I wouldn't think it is pleasant}" (G4P3).

Three groups disliked the mitigation subspace as they indicated ``\textit{During a meeting, the window and also the option to look into the deep is necessary for me.}" (G1P1-MA1) while also ``\textit{It is nicer to have light}" (G3P2-MA2).
Two additional groups were critical of the intervention subspace as it completely blocked their physical access to the door: ``\textit{I think it is good [...] being a little bit more private and enclosed. But there should always be a little bit of doorway}"; ``\textit{They gave me a more closed feeling because I couldn't see the door to leave the room anymore}" (G3P1); ``\textit{[...] It is nice to see if people come inside the room, it feels more safe}" (G7P3). 

\subsubsection{Face-to-face Meeting}\label{sec:adaptationcollaborative}
Participants in F2F meetings rated the mitigation subspace as more sociable than the intervention subspace.
Eight groups felt comfortable with both adaptation strategies, although they slightly preferred the mitigation subspace because it was more spacious: ``\textit{When we are doing the task with two [...], I always book a very large room so we can walk. So doing a job with two of us in a space like [the intervention subspace], for me that doesn't work}" (G4P1), taking into account that some participants in the same group were not acquainted with each other (G5P3). 

\begin{figure*}[t]
    \centering
    \includegraphics[width=1\linewidth]{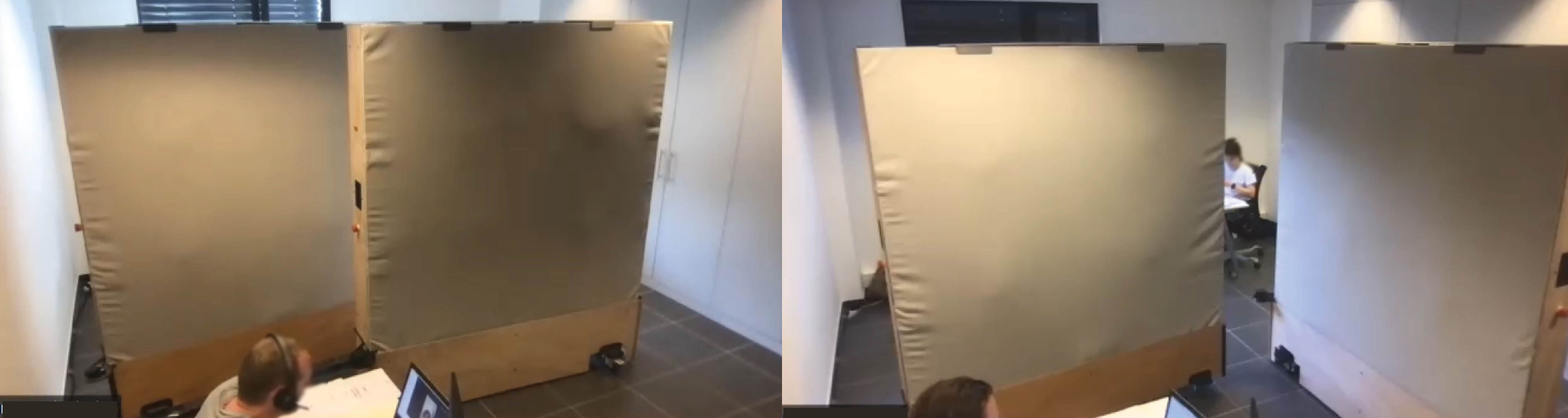}
    \caption{While engaged in a remote meeting, a P1 participant retained access to the exit and could exit the mitigation subspace (MA1-left); however, the same access was obstructed for another P1 participant by the intervention subspace (IA1-right).
    }
    \Description{Two photos of how the partitions affected participants who were in a remote meeting. (left) The partitions are positioned in the mitigation subspace (MA1), P1 who is in a remote meeting still has access to the exit and can leave the subspace. (right) The partitions are positioned in the intervention subspace (IA1), P1 who is in a remote meeting does not have access to the exit and cannot leave the subspace.}
    \label{fig:banner4.1.2}
\end{figure*}

\paragraph{Introducing Distractions}
One group did not understand why the intervention subspace for two participants in an F2F meeting was smaller than the mitigation subspace for one participant in a remote meeting: 
``\textit{... I felt like it (subspace) was the same size when I was by myself and during the meeting. Maybe it can be bigger when there are two people? It shouldn't be smaller}" (G7P1).

One group emphasised that participants in an F2F meeting should always be able to walk back to their desks instead of waiting for the partitions to open up, either to conclude a meeting on their own (G7P3) or to remedy an unforeseen issue: ``\textit{I am thinking purely practical because she didn't bring her laptop [...]. In this case, she should have easily gone back and come back because the [partitions] kind of combined both of the [desks] rather than isolating [manifesting an intervention subspace] us.}" (G7P1).

\subsection{Nudging Strategies}
Based on the two spontaneous meeting scenarios, this section addresses how participants interpreted the intentions of the nudging strategies, and how these inadvertently introduced new distractions.

\subsubsection{Face-to-face Meeting}\label{sec:nudgingcollaborative} 
While the spatial nudging strategy made the intentions to initiate or conclude a face-to-face meeting relatively clear, it offered participants no direct agency in directing its execution.

\paragraph{Interpreting the Nudging.} 
The intention of the spatial nudging strategy to initiate an F2F meeting was accurately interpreted by 15 out of 18 participants (\(83\%\)), based on four spatial affordances.
Two groups mentioned how the transient subspace regulated \textbf{visual access}, influencing whom to meet: ``\textit{It was quite clear because I could see [the participant I had collaborated with]}" (G3P2).
Conversely, one group noted how the transient subspace obstructed their visual access, which they linked to social unavailability: ``\textit{He was kind of blocked out, so I knew I had to go}" (G8P1).
Nine groups mentioned how the transient subspace regulated physical \textbf{accessibility}: ``\textit{It felt open enough to just [allow] me [to pass through] [...] I felt like it was good enough to go}" (G3P1).
Four groups interpreted the \textbf{directionality} of the transient subspace as `pointing' toward whom to meet: ``\textit{It [subspace] was pointed towards [another participant]}" (G4P1); ``\textit{The [partition] in front of me stayed not perpendicular to the wall but a little bit angled so that it directed me to [Participant 1]}" (G7P2).
Lastly, one group relied on \textbf{continuity} of movement to identify the right moment: ``\textit{I just waited until they didn't move [...] And then thought 'oh, it's ok now'}" (G6P3).
In 4 out of 18 instances (\(22\%\)), participants initiated an F2F meeting at an unintended moment. As shown in \autoref{fig:banner4.2}, in two instances (\(11\%\)) two participants simultaneously initiated an F2F with a third colleague. These misinterpretations occurred particularly when multiple spatial affordances were perceived to conflict with each other:  
``\textit{I think it was confusing at the beginning because I could still see [another participant I had to collaborate with]. But it wasn't my time, so I was told to go back}" (G7P3). 
Likewise, the intention to conclude a spontaneous meeting by opening up the robotic partitions to manifest the same transient subspace was correctly interpreted by 15 out of 18 participants (\(83\%\)), because participants became accustomed to its implied meaning.

\begin{figure*}[t]
    \centering
    \includegraphics[width=1\linewidth]{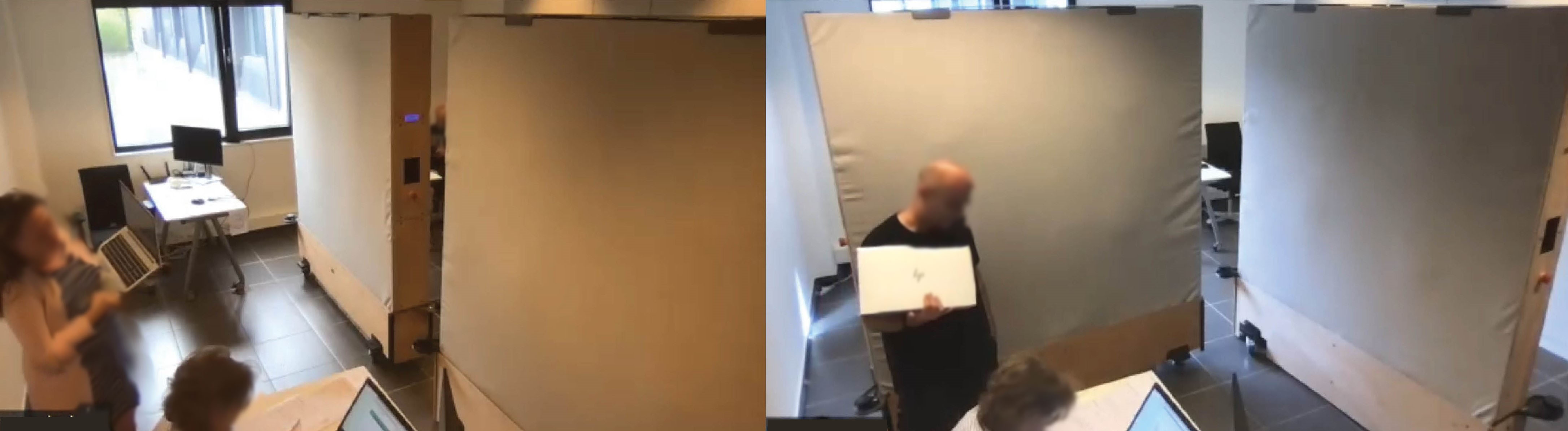}
    \caption{During an F2F meeting, the mitigation subspace (MA2-left) enabled both participants (P1 and P3) to access their desks, whereas an intervention subspace (IA1-right) blocked access for one participant (P2).}
    \Description{Two photos of how the partitions manifested a collaboration subspace. (left) The partitions create the mitigation subspace (MA2) for P1 and P3, whereby they have access to both desks. (right) The partitions create the intervention subspace (IA1) for P1 and P2, whereby P2 does not have access to his own desk anymore.}
    \label{fig:banner4.1.3}
\end{figure*}

\paragraph{Introducing Distractions.}
However, participants felt uncomfortable when the robotic partitions managed the initiation or conclusion of a spontaneous face-to-face meeting. While this sentiment was less apparent when participants were eagerly anticipating meeting their colleague, it became more pronounced when the partitions aimed to conclude an ongoing F2F meeting: ``\textit{The [partition] itself was very clear for me when I had to leave. Like they suddenly open, like, okay `get out'}" (G1P3). 
In six out of 18 collaborations (\(33\%\)), participants ignored the nudge to conclude their F2F meeting by the partitions opening up,  choosing instead to wait until their conversation ended naturally: ``\textit{We weren't finished completely yet [...]. So I just waited until she stopped writing. I also said that `it looks like the [partitions] want us to stop'}" (G8P1). 
In two cases, the spontaneous meeting was concluded before the partitions began moving because their collaboration task was accomplished prematurely. As participants began to move back to their desks, the partitions also started to open up to accommodate the conclusion of the meeting.

Four participant groups expressed a perceived lack of agency when they felt obligated to initiate or conclude their spontaneous F2F meeting: ``\textit{[During] collaboration, they move open to invite me to go back to my space. I find it, as I said, counter-intuitive because the collaboration isn't really over [...]}" (G6P1); ``\textit{It is strange. It is not common that a [partition] tells you what to do}" (G8P2); ``\textit{It is just more like you start collaborating and it locks you in and it doesn't tell you it is going to do that}" (G7P1). 

\subsubsection{Remote Meeting}\label{sec:nudgingindividual}
The gestural nudging strategy, aimed at guiding a participant into a subspace dedicated to a remote meeting, proved distracting due to the visual and auditory disruptions caused by the robotic motions and the participant moving around.

\paragraph{Interpreting the Nudging.} 
Although participants were intentionally familiarised with the gestural motions of the robotic partitions beforehand, four groups failed to recognise their gestural nudging when it occurred: ``\textit{Why are you [robotic partition] doing that? I think I already forgot that I had to move}" (G8P1).
Three groups did not notice the gestural motion: ``\textit{I didn't notice. I was talking to [the researcher]}" (G5P3). Even when they understood the gestural motion, they could not interpret which subspace they should move to: ``\textit{But still when it is moving, you still don't know by heart where you should go. It is not telling you that}" (G2P2). 

\begin{figure*}[t]
    \centering
    \includegraphics[width=1\linewidth]{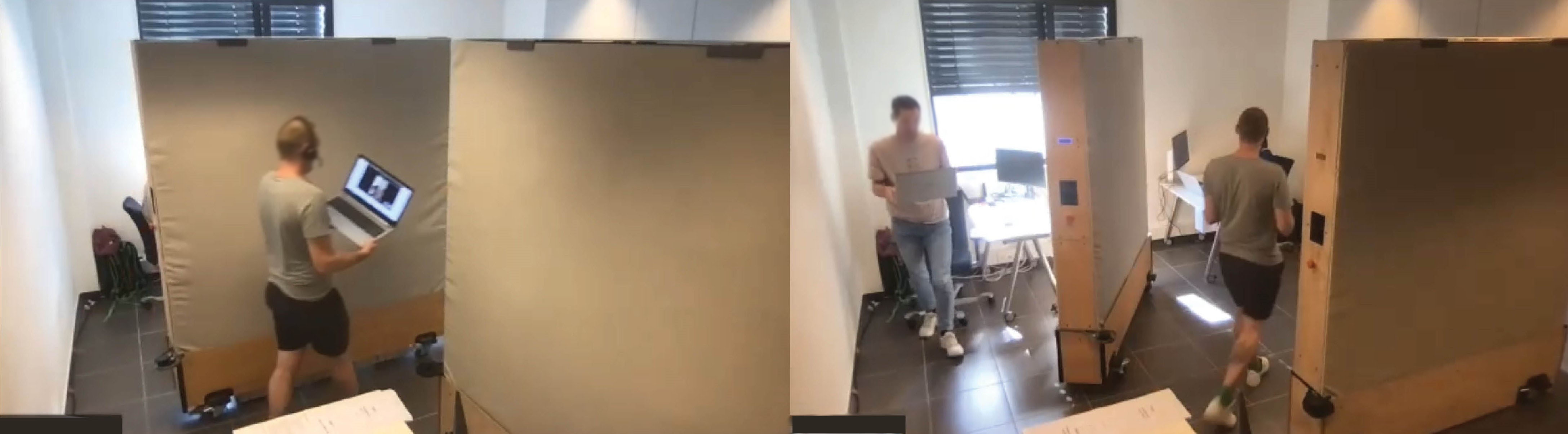}
    \caption{(Left) P1 was gesturally nudged (GN1) to relocate during a remote meeting. (Right) P1 was spatially nudged (SN1-2) to initiate an F2F meeting with P2, which yet inadvertently also nudged P3 to approach P2 at the same time.}
    \Description{Two photos of how the partitions nudged participants to another subspace. (left) P1 is gesturally nudged (GN1) during the remote meeting and stands up. (right) Participant 1 is spatially nudged (SN1-2) to have a physical meeting with P2, but P3 also stands up to approach P2 at the same time.}
    \label{fig:banner4.2}
\end{figure*}

After understanding the intention of gestural nudging, three participants recognised its benefit and agreed to follow its intention: ``\textit{I like to get up and go to some place because it takes you from one task to another mentally}" (G3P1); ``\textit{For the movement [physical activity], it was nice to go over there}" (G1P2), as shown in \autoref{fig:banner4.2}.

\paragraph{Reducing Distractions.}
Two groups felt less distracted when the partitions kept on covering the distracting participant by following their trajectory to the intervention subspace: ``\textit{I kind of noticed that he was walking. So when I heard the footsteps, I turned. But then I realised that he was covered with the [partitions] [...]. I guess it helped me to focus}" (G7P3).

\paragraph{Introducing Distractions.}
Seven groups found the `swaying' motion of the gestural nudging visually \textit{[...] was distracting and disturbing}" (G8P3), particularly when the target participant did not enter the intended subspace in a timely manner due to misunderstanding it: ``\textit{I was distracted by the dancing, but just because I wanted to tell him [the participant being nudged]}" (G8P1).
In cases where the target participant followed the nudging, the other participants felt distracted by their visibility: ``\textit{The moving of the [partition] distracted me as well as my colleague calling and moving around. There was too much movement}" (G3P3), and their noise: ``\textit{When the [partition] is moving, my colleague also moved around which causes lots of noise distracting me}" (G7P2). 

\subsection{Robotic Partitions}\label{sec:roboticresults}
Despite the initial familiarisation phase, eight out of nine participant groups still intentionally observed the motions of the robotic partitions: ``\textit{I was curious if [the partitions] were going to move and I was just waiting like when they are going.}" (G8P2).
This novelty effect was particularly apparent when partitions moved closely to the participants: ``\textit{[The partition] was moving [away from me]. So I felt less distracted}" (G1P3), ``\textit{The moving [partition] is far away from me, which distracts me less}" (G7P2). 
Two groups considered the speed of the partitions to be sufficiently safe and unobtrusive: ``\textit{It doesn't feel unsafe for me, because they are slow and steady}" (G8P3), ``\textit{[The speed] was okay, I think. If it was just one smooth movement, then I didn't notice it moving}" (G3P3). However, two groups found their slowness to be distracting:
``\textit{They move so slowly [...] then you would say 'never mind'}" (G8P3); ``\textit{ I don't know that when I have a meeting with someone that I would wait for the [partitions]}" (G3P3).
Once manifesting a static subspace, five groups found the robotic partitions to be unobtrusive: ``\textit{Once they stand still, you no longer pay attention to them...}" (G8P3), because ``\textit{After a while, I realised that I didn't notice them}" (G2P3) or because ``textit{[...]I was on a call and not doing very focused work, it did not affect me very much}" (G3P2).
As shown in \autoref{fig:banner4.1.1}, one group was distracted during their focused work by a manifested subspace because the partitions were misaligned with the orthogonality of the room:
``\textit{[...] I was constantly thinking 'Why are [the partitions] standing like this?'. They need to be straight. I wanted to stand up and move them. So that was a bit distracting [...]}" (G3P3). 
This partition misalignment also caused one group to misunderstand the intervention subspace: 
``\textit{They were in weird corners. They were not putting a cocoon around her or not really on a straight line}" (G2P2-IA2).


\section{Discussion}
Our findings reveal how the adaptation and nudging strategies were able to reduce the prototypical acoustic, visual, and privacy disturbances of a spontaneous meeting to some extent, while yet introducing various new disturbances. Our discussion thus analyses when and how a subspace should be manifested.  

\subsection{Managing Expectations}
Our results show that participants interpreted a robotically manifested subspace according to their personal expectations, which are shaped by the morphology, dynamic motion, and static alignment of the robotic partitions. This highlights the importance of managing human expectations through intentional robotic design.

\subsubsection{Morphological Expectations}
Participants felt distracted by how the morphological appearance of commercial-grade acoustic partitions contradicted their first-hand experience of perceiving no improvement in acoustic performance (see \nameref{sec:adaptationobtrusive}). 
This lack was caused by reasons outside of the robotic control, such as the sound-reflecting workspace surfaces that allowed the distracting sounds to bounce off the ceiling and walls.  
It nonetheless also highlights how the physical morphology of even a basic robotic furniture element like a partition should not over-promise its true capabilities \cite{cha2015perceived}. As the limited acoustic performance even reduced trust in its visual and privacy capabilities, it shows how one limitation can also negatively influence its other capabilities. 

\subsubsection{Expectations of Robotic Behaviour}
Participants were distracted by how the motion patterns of the robotic partitions to manoeuvre relatively heavy acoustic panels in a small space mismatched their expectations of how an efficient robot should behave. The participants were distracted when the robotic partitions exhibited erratic motions to avoid obstacles like sharp turns, abrupt manoeuvres, or sudden stops in their close proximity (see \nameref{sec:nudgingindividual}). 
As human visual perception is highly sensitive to even small dynamic changes in the peripheral field of vision \cite{faraday1997designing, bartram2003moticons}, a moving element of architectural scale inevitably captures visual attention. 
As humans expect mobile robots to behave in predictable \cite{sirkin2015mechanical}, efficient, and even purposeful \cite{ray2008people} ways, perceiving erratic motions over a considerable amount of time (approximately 30\% of our study) probably worsened the distraction. 
As such, this issue could be ameliorated once the orchestrated movement of multiple robotic partitions of architectural scale in a complex human-inhabited environment can be technologically mastered.

\subsubsection{Expectations of Architectural Behaviour}
Participants were distracted by how the misalignments of the arrangement of a subspace mismatched their expectations for architectural precision.
This misalignment mainly arose when adjoining partitions did not properly orthogonally align, or left small gaps between them. 
Not only did this misalignment undermine the expectation of a potentially seamless and intricate architectural appearance (see \nameref{sec:adaptationfocused}), it also negatively impacted the acoustic and privacy performance as distracting sounds and views could more easily escape the subspace.  
This issue aligns with findings in environmental psychology, where architectural space is typically processed in the background of people's attention, but quickly becomes a focal concern when they perceive even minor misalignments that conflict with their familiar expectations \cite{rooney2017}.
This issue could partly be attributed to how our recruitment and familiarisation process may have unintentionally over-promised the futuristic robotic capabilities to the participants, and how the relatively short study scenarios did not provide sufficient time for participants to learn the true capabilities of the partitions \cite{muir1996trust}.


\aptLtoX{\begin{shaded}
\noindent We suggest that purposeful robotic furniture such as office partitions should intentionally manage human expectations by design, which should ensure that its physical morphology (e.g. materials, form) aligns with its actual capabilities; its motion performs in a deliberate and predictable way; and its architectural arrangement manifests with spatial precision.
\end{shaded}}{\begin{tcolorbox}
We suggest that purposeful robotic furniture such as office partitions should intentionally manage human expectations by design, which should ensure that its physical morphology (e.g. materials, form) aligns with its actual capabilities; its motion performs in a deliberate and predictable way; and its architectural arrangement manifests with spatial precision.
\end{tcolorbox}}

\subsection{Balancing Distractions}
Our results show that participants perceived the trade-offs between reducing existing distractions and introducing new distractions were shaped by their work activities, location, and the duration of exposure. This suggests that such trade-offs could potentially be balanced at both the subspace and overall workspace levels.

\subsubsection{Balancing Subspace Distractions}
Contradictorily, participants engaged in remote meetings preferred the intervention strategy, while those engaged in focused work favoured the mitigation strategy, primarily because they perceived these respective subspaces as better suited to support their specific \textbf{work activities}.

As revealed in \nameref{sec:adaptationobtrusive}, participants engaged in remote meetings were more willing to accept the obstructed views and limited space of the more enclosed intervention subspace, as they perceived it to provide a sense of privacy and cosiness that is well-suited for holding remote meetings.
In contrast, \nameref{sec:adaptationfocused} revealed how participants engaged in focused work felt more disturbed by the distractions caused by the `leftover' subspace from the same adaptation strategy, as it lacked an outdoor view and social connectivity, which they considered essential for conducting focused work. 
Because the mitigation subspace did not have these shortcomings and even allowed them to share the same subspace with a colleague engaged in an identical work activity, they considered this strategy more suitable for conducting focused work. 
This contradiction could be explained by how a focused worker is more observant and sensitive to their immediate surroundings, compared to a worker who is viscerally engaged in an audiovisual conversation. This perspective also aligns with insights from environmental and social psychology, which indicate that workers prefer environments specifically tailored to support the tasks or activities most important to them \cite{schweiker2016effect}, and even overlook certain distractions in favour of the overall benefits that these environments provide \cite{Weber2022}. Workers also prefer sharing the same space when performing similar tasks, which fosters a sense of community and social cohesion \cite{lee2005effects}.

The newly introduced distractions were more noticeable among participants who had previously benefited from certain \textbf{spatial privileges}, compared to those who had never experienced them. Participants who were distracted by how a subspace obstructed their views towards the window, doorway, or colleagues engaged in the same work activity were occupying an office desk that originally provided these spatial qualities. 
Similarly, participants who were enjoying such spatial qualities were more resistant to being nudged towards entering a subspace that lacked these qualities. As revealed by \nameref{sec:nudgingindividual}, participants benefiting from natural light, for instance, were reluctant to relocate to an enclosed intervention subspace with less natural light.
This phenomenon highlights the importance of territoriality, as workers resist losing previously enjoyed privileges \cite{Kahneman1991}, including their agency to avoid distractions in a given space and the psychological benefit of feeling in control over their surroundings \cite{brown2009claiming}.

Furthermore, participants engaged in remote meetings reported that, as the \textbf{duration} of the meeting increased, the distractions they were initially willing to accept became more noticeable, shifting their preference from an intervention to a mitigation subspace (see \nameref{sec:adaptationobtrusive}). In terms of the manifestation process of a subspace, we similarly noticed how participants located further away from unintended distractions in terms of \textbf{distance} were less affected by them.

\subsubsection{Balancing Workspace Distractions}
Because participants perceived the trade-off between reduced and newly introduced distractions differently depending on their work activity, a challenging conflict arises: when workers engage in different work activities in the same workspace, the ideal subspaces to support them cannot physically coexist. 

The realisation that each individual trade-off is influenced by \textbf{work activity, spatial privileges, exposure duration, and distance}, suggests that this conflict could potentially be resolved by balancing the trade-offs of each individual worker in the same workspace in a more holistic manner. 

We thus propose that a future autonomous robotic partition system could model coexisting individual trade-offs as heuristic measures that are dynamically weighted according to the situation at hand. For instance, the selection of a nudging or adaptation strategy could prioritise reducing distractions for workers engaged in longer-duration work activities, minimising the risk that they would eventually become distracted over time by matters like lack of privacy, outdoor view or social connectivity. 
Similarly, an adaptation strategy could exempt existing experiences of certain spatial privileges, or vice versa, to ensure that all workers can enjoy the spatial privileges equally over one workday.
Finally, the gestural nudging strategy should only be manifested at a sufficient distance from focused workers, whereas the spatial nudging strategy should only be considered when all workers engaged in focused work are located in the vicinity of the source of distraction and vulnerable to being distracted.


\aptLtoX{\begin{shaded}
\noindent We propose that robotic furniture like office partitions can equitably balance its purposefully introduced support and inadvertently-introduced distractions by distributing its manifestation based on the relative activities, distances, spatial privileges, and exposure durations of co-located individuals.
\end{shaded}}{\begin{tcolorbox}
We propose that robotic furniture like office partitions can equitably balance its purposefully introduced support and inadvertently-introduced distractions by distributing its manifestation based on the relative activities, distances, spatial privileges, and exposure durations of co-located individuals.
\end{tcolorbox}}

\subsection{Conveying Intent}
Our results show that participants found an adaptation or nudging strategy more understandable when it interacted with them indirectly through the spatial affordances of the transiently manifested subspace, as opposed to the direct gestural motions.

\subsubsection{Indirect Interaction}
Even when the robotic partitions were manually wizarded to appear fully autonomous, the participants still interacted with them indirectly by observing the architectural affordances of the manifested subspaces. 
The architectural affordances, like surface area, directionality, visibility regulation, and accessibility served as interpretable cues to denote the intention of the subspace, which caused participants to react accordingly.
For instance, participants in remote meetings understood that the tight enclosure of the intervention subspace was meant to provide acoustic and visual privacy, causing them to talk more loudly and even gesture more extensively.
As revealed in \nameref{sec:nudgingcollaborative}, participants interpreted the directionality of the partitions as pointing towards specific subspaces or colleagues, and understood that their physical manifestations regulated their view towards colleagues as either inviting or blocking potential meetings.  
Given that humans can perceive the purpose of an architectural space based on its visual consistency \cite{Wicker2002}, relative scale \cite{Bolbroe2016}, or the affordances offered by the physical environment towards ongoing activities, our findings add the notion that robotic furniture can convey its intentions in an architectural way.
As such, the area, contour or directionality of a robotically manifested subspace has the potential to invite workers to initiate an individual or collaborative work activity, nudge them to relocate before or during a work activity, ideally afford a focused or a meeting activity, or even nudge them to conclude their work activity. 
These findings can be partly explained by how humans perceive an architectural space through interpreting its spatial shape \cite{banaei2017walking}, to the extent that humans can use this shape to navigate in complex buildings \cite{dalton2014navigating}. 
While we know that such perceptual processes rapidly decline as people become more accustomed to spatial shapes \cite{banaei2017walking}, it is still unknown whether people are able and willing to interpret continuously and dynamically changing architectural spaces. 

\subsubsection{Direct Interaction}
Our results strongly suggest that direct interactions like our deliberate gestural motion pattern need to be sufficiently intuitive to be understood, while also being sufficiently unobtrusive for those who are not meant to engage with it.
For instance, although we had informed all participants about the actionable meaning of the `swaying' motion as an emblematic gesture \cite{dave2023} to enter an adjoining subspace, it still regularly failed to guide the intended participant (see \nameref{sec:nudgingindividual}), and distracted all other participants. 
These findings align with research in HRI that indicates how the effectiveness of robotic nudges depends on the familiarity  \cite{gronvall2014causing} and past personal experiences \cite{Washburn2020} of users. 
This issue could be solved by translating the intention of a nudge into a natural affordance of the robot, which is also reaffirmed by how our spatial nudge was more intuitively understandable than the gestural nudge. 
As research in HRI also confirms, humans can rely on directional spatial cues when interpreting robotic furniture gestures, which can either be translated from how a piece of furniture is typically used in space, like the opening of a door \cite{ju2009approachability} or the adjustment of a chair \cite{Agnihotri2019}; or be directly related to human proxemic zones, like gradually approaching \cite{sirkin2015mechanical} or connecting them \cite{takashima2015movementable}.
Knowing that people feel more comfortable performing public tasks in group \cite{Zajonc1965}, the apparent success of the spatial compared to gestural nudging strategy could also be attributed to how two participants could visually signal and reassure each other through the transient subspace, as opposed to feeling socially embarrassed when misinterpreting the gestural intentions.


\aptLtoX{\begin{shaded}
\noindent We propose that purposeful robotic furniture like partitions can nudge co-located occupants unobtrusively and understandably by conveying its intentions through the architectural affordances that transiently regulate the area, directionality, and visibility of the space.
\end{shaded}}{\begin{tcolorbox}
We propose that purposeful robotic furniture like partitions can nudge co-located occupants unobtrusively and understandably by conveying its intentions through the architectural affordances that transiently regulate the area, directionality, and visibility of the space.
\end{tcolorbox}}


\section{Limitations}

Despite our relatively small \textit{participant cohort} (\(n=27\)), their expertise in working in open-plan office layouts and unfamiliarity with robotic technology offered invaluable ecologically valid insights into how robotic furniture can benefit everyday knowledge workers. The demographics of our participants, consisting predominantly of women (19 out of 27 participants, 70.4\%) with an average age of 35 years (\(SD=11.0\)), yet capture an under-represented demographic in robotic research \cite{Hopko2022}. However, unique subjective characteristics among our participants might affect the transferability of the results. For instance, one participant suffering from migraine noted that the shifting natural lighting caused by the moving partitions triggered headaches (G9P1), while two others explicitly expressed the ability to ignore distractions in an office environment (G8P1 and G5P1). Furthermore, four participants showed scepticism towards the concept of robotic furniture during the experiment, which might have affected their experience. 

The \textit{semi-control methodology} may have influenced the ecological validity of our findings. By simulating the conflicting work activities within a short duration (one hour), with only one participant engaging in a remote meeting, we did not capture the full complexity of a typical open-plan office workday. 
In real-world scenarios, such settings often involve workers undertaking a broad spectrum of tasks, including multiple simultaneous planned and spontaneous meetings. Additionally, the short duration of each study, chosen to accommodate the busy schedules of the participants, limited their exposure to the robotic partitions, which in turn negatively impacted their trust.
The choice of group interviews, aimed at fostering discussions among participants, might have biased the feedback towards more outspoken or already acquainted groups of participants.

Lastly, the somewhat atypical \textit{desk layout typology}, characterised by the empty space between the work desks needed to allow the partitions to manoeuvre around, limits the transferability of our findings.


\section{Conclusion}
In this study, we investigated how two autonomous robotic partitions can synchronously manifest an enclosed subspace within an open-plan office layout to reduce the distractions originating from spontaneous face-to-face and remote meetings. We evaluated how nine groups of three collocated participants experienced two adaptation strategies aimed at enclosing the source of distraction or enclosing others affected by it, and two nudging strategies aimed at relocating participants having meetings to a more suitable subspace via motion-based gestures or transient subspaces. 

Based on our findings, we proposed three design considerations: robotic furniture should manage human expectations through intentional design by aligning its morphological, static and dynamic qualities with its actual capabilities; it should balance distractions equitably among co-located people by considering their activity, distance, spatial privileges, and exposure durations; and it could nudge people subtly and intuitively by leveraging the architectural affordances of transient subspaces that regulates visual and physical access among them.
These insights potentially can inform the design and deployment of robotic furniture as well as any other mobile robots in human-centred environments. By utilising architectural affordances, such systems can potentially enhance their ability to reduce temporal distractions, support human activities and therefore promote longer-term environmental satisfaction, health, and well-being.


\begin{acks}
This research is supported by the KU Leuven ID-N project IDN/22/003 ``Adaptive Architecture: the Robotic Orchestration of a Healthy Workplace".
We kindly thank IDEWE for allowing us to conduct this study at their office and for their support in managing the practicalities of the study deployment. We also thank prof. Lode Godderis (Centre for Environment and Health, Department of Public Health and Primary Care, KU Leuven) and prof. Herman Bruyninckx (Robotics, Automation and Mechatronics, Department of Mechanical Engineering, KU Leuven) for their valuable advice on the study design. Finally, we thank all participants for taking the time from their busy schedules to attend and contribute to our research.
\end{acks}

\bibliographystyle{ACM-Reference-Format}
\balance
\bibliography{manuscript}

\end{document}